\documentclass[11pt,a4paper]{article}
\usepackage{graphicx}
\makeatletter
\def\section{%
    \@addtoreset{equation}{section}
    \def\theequation{\thesection.\arabic{equation}}
    \@startsection {section}{1}{\z@}{-3.5ex plus-1ex minus
    -.2ex}{2.3ex plus.2ex}{\reset@font\large\bf}}
\makeatother
\newcommand{\lvm}{\leavevmode}

\newcommand{\del}{\partial}

\newcommand{\lsim}{\mbox{ \raisebox{-1.0ex}{$\stackrel{\textstyle <}
{\textstyle \sim}$ }}}
 
\begin{document}
\baselineskip 15pt
\begin{flushright}
     OCHA-PP-191
\end{flushright}

\medskip
\begin{center}
{\LARGE 
\bf
 Probing Noncommutative Space-Time in the Laboratory Frame
}
\\

\bigskip
{\Large Jun-ichi Kamoshita}
\\

{\it 
   Institute of Humanities and Sciences, Ochanomizu University,\\
and\\
   Department of Physics, Ochanomizu University,\\
     2-1-1 Otsuka, Bunkyo, Tokyo 112, JAPAN
 }

\bigskip
{\bf Abstract}

\medskip
\begin{minipage}{140mm}
{ The phenomenological investigation
 of noncommutative space-time
 in the laboratory frame are presented. 
 We formulate
 the apparent time variation of
 noncommutativity parameter $\theta_{\mu\nu}$
 in the laboratory frame
 due to the earth's rotation.
 Furthermore, in the noncommutative QED,
 we discuss how to probe the electric-like component
 $\overrightarrow{\theta_{E}}=(\theta_{01},\theta_{02},\theta_{03})$
 by the process $e^-e^+\to\gamma\gamma$ 
 at future $e^-e^+$ linear collider. 
 We may determine
 the magnitude and the direction of $\overrightarrow{\theta_{E}}$ 
 by detailed study of 
 the apparent time variation of total cross section.
 In case of us observing no signal,
 the upper limit on the magnitude of
 $\overrightarrow{\theta_E^{}}$ 
 can be determined
 independently of its direction.
}

\end{minipage}
\end{center}

\medskip
\section{Introduction}
 The early study of noncommutative space-time was presented
 by Snyder\cite{Snyder} in 1947,
 with respect to the need to regularize
 the divergence of quantum field theory.
 In Snyder's work, it was suggested that
 the divergence may be regularized by
 an elementary unit of length
 induced by the noncommutativity of space-time.
 Snyder's basic idea was 
 the extension of the quantization of phase space
 in quantum mechanics.
 Furthermore, noncommutativity of space-time
 may arise from string theory in 
 the specific low energy
 limit\cite{Connes}.
 The noncommutative space-time is characterized by 
 operators $\hat{X}_{\mu}$ satisfying 
 the commutation relation
\begin{eqnarray}
  [\hat{X}_\mu,\hat{X}_\nu]=i\theta_{\mu\nu}
\label{eqn:XhatNC}
\end{eqnarray}
 where $\theta_{\mu\nu}$ is antisymmetric constant matrix,
 $\theta_{\mu\nu}=-\theta_{\nu\mu}$ and 
 $[\hat{X}_\rho,\theta_{\mu\nu}]=0$.
 And $\theta_{\mu\nu}$ have dimension of (Length)${}^{2}$. 
 Therefore, (\ref{eqn:XhatNC}) introduce
 the elementary unit of length in the theory,
 such as Planck constant in quantum mechanics.
 Nonzero constant matrix $\theta_{\mu\nu}$ may
 violates Lorentz invariance.
 Lorentz violation in noncommutative quantum field theory
 have been studied\cite{LVNCFT}.

 It is known that QED in noncommutative space time
 (NCQED)\cite{NCQED} is invariant under 
 the noncommutative version of $U(1)$ gauge transformation and
 is renormalizable
 at one loop level\cite{NCQED,Hayakawa,anomaly}.
 Axial anomaly\cite{anomaly} and CPT invariance\cite{CPT}
 in NCQED have also been studied.
 There are several phenomenological study on NCQED
 for low energy experiments\cite{gmu,lowEXP,LambShift,ABeff}. 
 Assuming $\theta_{\mu\nu}$ is constant in the laboratory frame,
 a lower bound on noncommutativity scale $\Lambda_{NC}$ 
 have been found to be $\Lambda_{NC}>100$GeV\cite{LambShift} 
 in order that the result of Lamb shift is consistent with
 the ordinary quantum mechanics.
 Other limit on noncommutativity parameter have been found to be
 $\theta\lsim{(10\mbox{TeV})^{-2}}$, 
 if $\theta_{\mu\nu}\equiv\theta\epsilon_{\mu\nu}$, 
 by an analysis of noncommutative Aharonov-Bohm effect\cite{ABeff}.
 High energy phenomenology in NCQED has also been studied 
 for several processes at future linear
 colliders\cite{Hewett,NCQEDNLC}.
 Moreover, phenomenology relevant to
 Standard Model (SM) like interactions in noncommutative space-time
 have also been studied \cite{NCSM,NCHiggs}
 on the assumption that we may obtain
 SM-like interaction in noncommutative space-time
 by usual procedure replacing every products of fields
 with the star product.
 In those previous studies,
 however, the direction of $\theta_{\mu\nu}$ have been 
 assumed to be fixed to the laboratory frame.
 Such an assumption might be justified, 
 if measurements would be given by 
 the data set suitably averaged over time
 and also over polar angle distributions.
 
 The $\theta_{\mu\nu}$, however, may be considered 
 as an elementary constant in the nature.
 And there may exist a class of specific coordinate system
 in which the direction of $\theta_{\mu\nu}$ is fixed.
 It is likely that such a coordinate system is
 fixed to the celestial sphere.

 On the contrary,
 the laboratory frame is located on the earth
 and is moving by the earth's rotation.
 Therefore, as was mentioned in\cite{Hewett,NCHiggs,NCK},
 we should take into account the apparent time variation of
 $\theta_{\mu\nu}$ in the laboratory frame
 due to the earth's rotation
 when we discuss phenomenology for 
 any experiment on the earth.
 In this paper,
 we will consider the effect of
 apparent time variation of $\theta_{\mu\nu}$
 in the experiments due to the earth's rotation seriously.

 If an anisotropy due to noncommutativity of space-time exists, 
 probing the specific direction of $\theta_{\mu\nu}$ and 
 measuring the magnitude of elementary unit of length
 are very interesting tasks from 
 both experimental and theoretical aspects.
 We may determine the direction of $\theta_{\mu\nu}$
 by the analysis taking into account 
 effects of time variation of the measurements.

 This paper is organized as follows.
 In section 2, we present the parameterization for 
 $\theta_{\mu\nu}$ including the effect of the earth's rotation.
 In section 3, we make some comments on
 the time dependent cross section and
 we define the time averaged cross section.
 In section 4, we show several numerical results and 
 we discuss how to prob $\theta_{\mu\nu}$
 at future linear collider experiments.
 Finally, we conclude our result and discussion. 

\section{Expression of ${\theta_{\mu\nu}}$ in the laboratory frame}
 The noncommutativity parameter $\theta_{\mu\nu}$
 can be classified into two parts. 
 One is the electric-like component 
 $\overrightarrow{\theta_E}=(\theta_{01},\theta_{02},\theta_{03})$.
 Another is the magnetic-like component 
 $\overrightarrow{\theta_B}=(\theta_{23},\theta_{31},\theta_{12})$.
 Those elements can be determined when
 a coordinate system is chosen.
 In the specific coordinate system,
 both $\overrightarrow{\theta_E}$ and $\overrightarrow{\theta_B}$ 
 should be constant vectors.
 Hereafter we call such a coordinate system
 a ``primary'' coordinate system.
 It is feasible that 
 we take a set of coordinates fixed to 
 the rest frame of the cosmic microwave background(CMB)
 as a ``primary'' coordinate system.
 According to COBE experiment \cite{COBE},
 the boost of the solar system for the CMB rest frame is
 about 370km/s. 
 This is about 0.12\% of the speed of light in vaccum.
 Moreover the speed of the earth in solar system is
 about 29.78km/s.
 Therefore the effect of the boost to the measurement of 
 $\overrightarrow{\theta_E}$ and $\overrightarrow{\theta_B}$ 
 are small enough to neglect
 in comparison with the detector resolution in
 the collider experiments.
 And we may consider that 
 the CMB rest frame is fixed to the celestial sphere
 approximately. Thus, hereafter, we assume that 
 a primary coordinate system and also each direction of 
 $\overrightarrow{\theta_E}$ and 
 $\overrightarrow{\theta_B}$
 are fixed to the celestial sphere effectively.

\begin{figure}[t]
\begin{center}\lvm
\begin{minipage}[c]{160mm}
\begin{tabular}{cc}
\begin{minipage}[c]{75mm}
\centerline{
\includegraphics[width=65mm, angle=0]{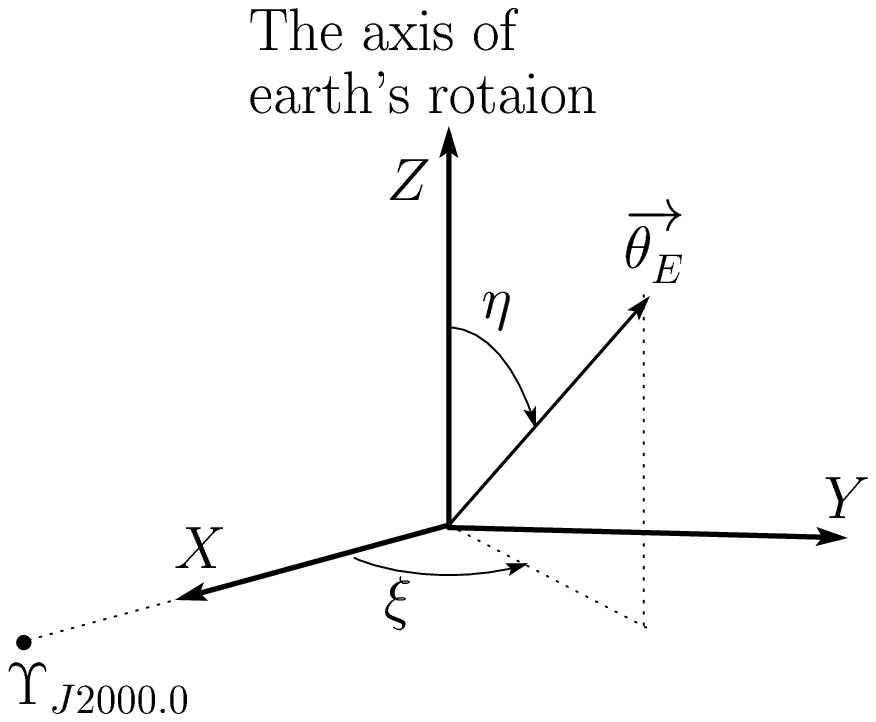}
}
\caption[]{ ``primary'' coordinate system ($X$-$Y$-$Z$).
      The axis  $X$ point to 
      the vernal equinox $\Upsilon_{J2000.0}$.
      The electric-like component 
      $\overrightarrow{\theta_E}$ of $\theta_{\mu\nu}$ is also shown.
      The direction of $\overrightarrow{\theta_E}$ is 
      parameterized by constant angle parameters $\eta$ and $\xi$.
        }
\label{fig:coordinates}
\end{minipage}
&\quad
\begin{minipage}[c]{75mm}
\centerline{
\includegraphics[width=65mm, angle=0]{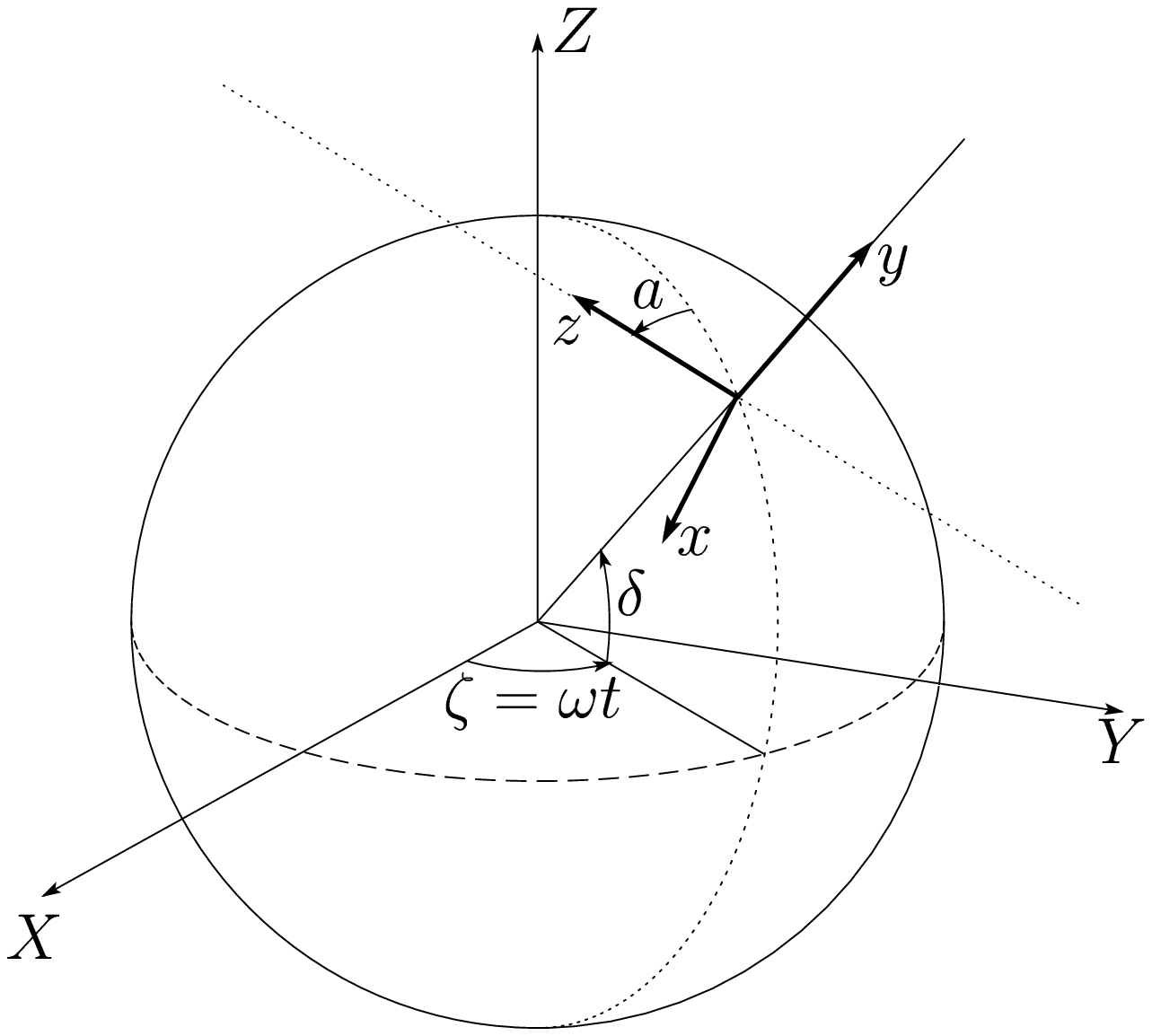}
}
\caption[]{Arrangement of laboratory coordinate system
     ($x$-$y$-$z$)
     for an experiment on the earth in 
     the ``primary'' coordinate system ($X$-$Y$-$Z$).
     $\delta$, $a$ and $\omega$ are constants.}
\label{fig:localframe}
\end{minipage}
\end{tabular}
\end{minipage}
\end{center}
\end{figure}

 At first, we introduce 
 a primary coordinate system which is Cartesian coordinate system.
 The $Z$ axis is along the axis of the earth's rotation and
 the positive direction of $Z$ axis points to the north.
 The axis pointed to the vernal equinox ($\Upsilon_{J2000.0}$)
 is labeled $X$.
 We take $X$-$Y$-$Z$ system as the right-handed system.
 Figure \ref{fig:coordinates} shows 
 the sketches of the primary coordinate system and 
 the direction 
 $\overrightarrow{\theta_E}$ 
 parametrized by $\eta$ and $\xi$.
%

 Let $\overrightarrow{e_X}$, $\overrightarrow{e_Y}$
 and $\overrightarrow{e_Z}$
 be the orthonormal basis of the 
 primary coordinate system ($X$-$Y$-$Z$).
 Then 
\begin{eqnarray}
  \overrightarrow{\theta_E}=\theta_E
  \left( \overrightarrow{e_X} \sin\eta\cos\xi 
        +\overrightarrow{e_Y} \sin\eta\sin\xi 
        +\overrightarrow{e_Z} \cos\eta 
  \right),
\label{eqn:th_prmry}
\end{eqnarray}
 where $0\le \eta\le\pi$, $0\le \xi\le 2\pi$ and
 $\theta_E^{}\equiv|\overrightarrow{\theta_E^{}}|$.
 To be exact, this $X$-$Y$-$Z$ coordinate system moves slightly
 owing to the earth's precession.
 Since the period of the earth's precession is
 about $2.6\times10^4$ years,
 the vernal equinox is moving by about 0.014 degree/year.
 Therefore, we can neglect the earth's precession
 during the term of experiments.

 On the other hand,
 the usual coordinate system for experiments
 is fixed to the detector. 
 We label each axis of such a coordinate system
 by small letter ($x$,$y$,$z$).
 As an example we consider an $e^-e^+$ collider experiment.
 The origin is set at the interaction point.
 The $z$ axis is along the direction of $e^-$ beam.
 The horizontal and vertical axes are
 labeled $x$ and $y$ respectively.
 The $x$-$y$-$z$ system should be the right-handed system.
 Hereafter we call this coordinate system 
 the ``laboratory'' coordinate system.

 As is shown in figure \ref{fig:localframe},
 we parametrize the location of an $e^-e^+$ experiment
 on the earth by a latitude $\delta$ of the detector site, 
 the angle $a$ between direction of $z$ axis and
 the meridian at detector site,
 and the angle $\zeta$ between $X$-$Z$ plane and $y$-$Z$ plane.
 The angle $a$ is measured counterclockwise from the north.
\footnote{Our definition of the angle $a$ is opposite to 
        the definition of usual azimuth in astronomy.
        We define angle $a$ as 
        it increases with 
        a positive rotation in the right-handed system.
        We may call the angle $a$ the counter-azimuth.
        } 

 Let $\overrightarrow{e_x}$, 
 $\overrightarrow{e_y}$ and $\overrightarrow{e_z}$
 be the orthonormal basis of the laboratory coordinate system
 ($x$-$y$-$z$).
 The transformation between 
 { ($\overrightarrow{e_X}$, 
    $\overrightarrow{e_Y}$, 
    $\overrightarrow{e_Z}$)  }
 and 
 { ($\overrightarrow{e_x}$, 
    $\overrightarrow{e_y}$, 
    $\overrightarrow{e_z}$) }
 is given by
\begin{eqnarray}
  \left(\begin{array}{c}
     \overrightarrow{e_X}\\ \overrightarrow{e_Y}\\ \overrightarrow{e_Z}
      \end{array}\right)
 &=& R 
   \left(\begin{array}{c}
     \overrightarrow{e_x}\\ \overrightarrow{e_y}\\ \overrightarrow{e_z}
   \end{array}\right),
\end{eqnarray}
\begin{eqnarray}
 R &=& \left(\begin{array}{ccc}
      c_\zeta &-s_\zeta & 0 \\
      s_\zeta & c_\zeta & 0 \\
      0 & 0 & 1
   \end{array}\right)
   \left(\begin{array}{ccc}
      c_\delta & 0 & -s_\delta \\
      0 & 1 & 0 \\
      s_\delta & 0 &  c_\delta 
   \end{array}\right)
   \left(\begin{array}{ccc}
      1 & 0 & 0 \\
      0 & c_a &-s_a \\
      0 & s_a & c_a 
   \end{array}\right)
   \left(\begin{array}{ccc}
      0 &  1 & 0 \\
     -1 &  0 & 0 \\
      0 &  0 & 1
   \end{array}\right)\cr
 &=&
   \left(\begin{array}{ccc}
      c_a s_\zeta + s_\delta s_a c_\zeta &  c_\delta c_\zeta 
                  &  s_a s_\zeta - s_\delta c_a c_\zeta \\
     -c_a c_\zeta + s_\delta s_a s_\zeta &  c_\delta s_\zeta 
                  & -s_a c_\zeta - s_\delta c_a s_\zeta \\
      -c_\delta s_a  &  s_\delta  &  c_\delta c_a
   \end{array}\right)
\end{eqnarray}
 with 
 $-\pi/2\le\delta\le\pi/2$ and $0\le{a}\le2\pi$,
 where we use the usual abbreviation, $c_\zeta=\cos\zeta$, etc.
 Hereafter we take the orthonormal basis of 
 the laboratory coordinate system as the usual way,
 $\overrightarrow{e_x}=(1,0,0)^T$, 
 $\overrightarrow{e_y}=(0,1,0)^T$ and
 $\overrightarrow{e_z}=(0,0,1)^T$. 
 Then, in the laboratory coordinate system,
 the orthonormal basis of 
 the primary coordinate system can be written as
\begin{eqnarray}
\begin{array}{ccc}
\overrightarrow{e_X}= 
   \left(\begin{array}{c}
      c_a s_\zeta + s_\delta s_a c_\zeta \\
      c_\delta c_\zeta \\
      s_a s_\zeta - s_\delta c_a c_\zeta 
   \end{array}\right),
&
\overrightarrow{e_Y}=
   \left(\begin{array}{c}
     -c_a c_\zeta + s_\delta s_a s_\zeta \\
      c_\delta s_\zeta \\
     -s_a c_\zeta - s_\delta c_a s_\zeta 
   \end{array}\right),
&
\overrightarrow{e_Z}=
   \left(\begin{array}{c}
      -c_\delta s_a  \\
       s_\delta  \\
       c_\delta c_a
   \end{array}\right).
\end{array}
\label{eqn:basis}
\end{eqnarray}
 Note that in the laboratory coordinate system
 the direction of $Z$ axis, 
 namely the axis of the earth's rotation,
 is given only by the location of $e^-e^+$ experiment
 ($\delta$, $a$). 
 For example, ($\delta$, $a$) of 
 LEP experiments \cite{kawamoto} are 
 approximately
 (46.15$^{\circ}$,  40$^{\circ}$) for OPAL, 
 (46.15$^{\circ}$, 130$^{\circ}$) for ALEPH, 
 (46.15$^{\circ}$, 220$^{\circ}$) for L3 and 
 (46.15$^{\circ}$, 310$^{\circ}$) for DELPHI.
 Therefore the arrangement of $\overrightarrow{e_Z^{}}$
 in the laboratory coordinate system
 of the LEP experiments are different 
 each other.
 
 The angle $\zeta$ increases with time $t$ 
 owing to the earth's rotation.
 A detector site will return to the same direction
 by a sidereal day, $T_{day}=$23h56m4.09053s \cite{PDG}.
 Therefore, we may take 
\begin{eqnarray}
  \zeta={\omega}t &{\rm with}& \omega\equiv{2\pi}/{T_{day}},
\end{eqnarray}
 by setting $t=0$ when the detector site is 
 on the $X$-$Z$ half plane with $X>0$.

 Substituting (\ref{eqn:basis}) into (\ref{eqn:th_prmry}),
 we find the expression of
 $\overrightarrow{\theta_E^{}}$
 in the laboratory coordinate system, 
\begin{eqnarray}
     \overrightarrow{\theta_E}
 &=& \overrightarrow{\theta_{EV}}
    +\overrightarrow{\theta_{ES}}, 
\label{eqn:thE}\\
 \overrightarrow{\theta_{EV}}&=&
 \theta_E\sin\eta
 \left[
 \left(\begin{array}{c}
   s_\delta s_a \\
   c_\delta     \\
  -s_\delta c_a
     \end{array}\right)c_{(\omega t-\xi)}
  +\left(\begin{array}{c}
   c_a \\  0   \\  s_a 
     \end{array}\right) s_{(\omega t-\xi)}
 \right],
\label{eqn:thEV}\\
 \overrightarrow{\theta_{ES}}&=&
 \theta_E\cos\eta
 \left(\begin{array}{c}
  -c_\delta s_a \\
   s_\delta     \\
   c_\delta c_a 
     \end{array}\right),
 \label{eqn:thES}
\end{eqnarray}
 where
 $\overrightarrow{\theta_{ES}}$ is 
 the projection of $\overrightarrow{\theta_E}$
 onto the $Z$ axis and is
 the stable part of $\overrightarrow{\theta_E^{}}$
 in the laboratory coordinate system.
 $\overrightarrow{\theta_{EV}}$ is
 the time variation part of $\overrightarrow{\theta_E}$.
 The direction of $\overrightarrow{\theta_{EV}}$ revolves on the
 $\overrightarrow{\theta_{ES}}$ by a period $T_{day}$. 
 This is the apparent time variation due to 
 the earth's rotation.
 Angle parameter $\xi$
 appears in the expression of $\overrightarrow{\theta_{EV}}$
 as the initial phase for time evolution.
 It is easy to show that
\begin{eqnarray}
\begin{array}{cc}
\left|\overrightarrow{\theta_{ES}}\right|
   =\left|\overrightarrow{\theta_E}\right|\cos\eta,&  
\left|\overrightarrow{\theta_{EV}}\right|
  =\left|\overrightarrow{\theta_E}\right|\sin\eta,\\ 
\overrightarrow{\theta_{ES}}\cdot\overrightarrow{\theta_{EV}}=0,&
  \left|\overrightarrow{\theta_{ES}}\right|^2
 +\left|\overrightarrow{\theta_{EV}}\right|^2
 =\left|\overrightarrow{\theta_{E}}\right|^2.
\end{array}
\end{eqnarray}
 Therefore the magnitude of each vector
 $\overrightarrow{\theta_{ES}}$,
 $\overrightarrow{\theta_{EV}}$ and
 $\overrightarrow{\theta_{E}}$
 is independent of time.

\begin{figure}[t]
\begin{center}
\begin{minipage}[c]{150mm}
 \begin{tabular}{cc}
 \begin{minipage}[c]{80mm}
 \centerline{
 \includegraphics[width=50mm, angle=0]{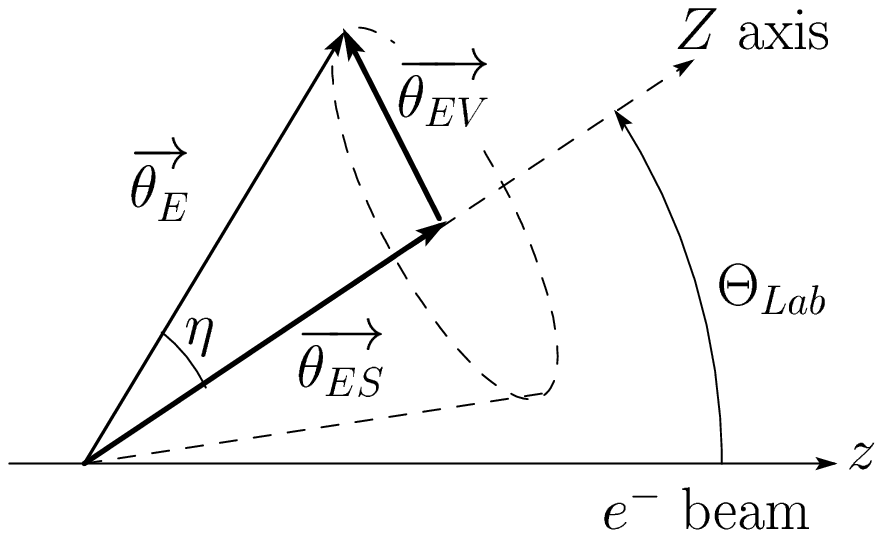}
 }
\vspace*{-1.3cm}
 (a) 
\end{minipage}
 &
\begin{minipage}[c]{80mm}
\centerline{
\includegraphics[width=50mm, angle=0]{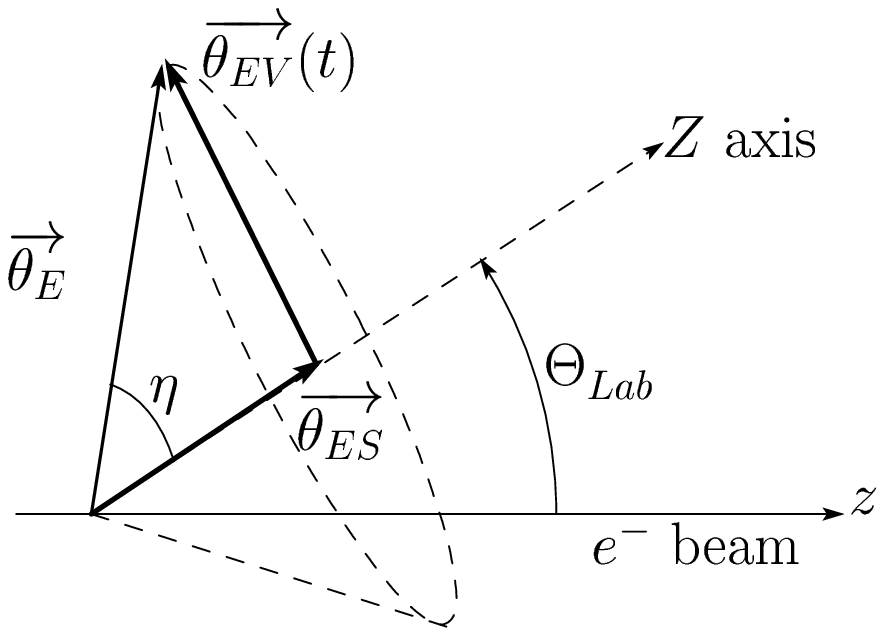}
}
\vspace*{-1.3cm}
(b)
\end{minipage}
\end{tabular}
\caption[]{Two typical time variation of $\overrightarrow{\theta_E}$ 
         in the laboratory frame.
  (a) for $\eta \le \Theta_{Lab}$ and
  (b) for $\eta \ge \Theta_{Lab}$,
  where $\cos\Theta_{Lab}={c_{\delta}^{}c_{a}^{}}.$
         }
\label{fig:thetaE}
\end{minipage}
\end{center}
\end{figure}

 Let $\Theta_{Lab}^{}$ be the polar
 angle of $\overrightarrow{\theta_{ES}}$ 
 in the laboratory coordinate system.
 From (\ref{eqn:thES}),
 we find 
$\cos\Theta_{Lab}^{}=c_{\delta}^{}c_{a}^{}. $
 We may classify the apparent time variation into
 two typical cases, $\eta\le\Theta_{Lab}^{}$
 and $\eta\ge\Theta_{Lab}^{}$.
 Figure \ref{fig:thetaE} shows the two cases for 
 the apparent time variation of
 $\overrightarrow{\theta_E}$ in the laboratory coordinate system.
 Let $\Phi_{E}^{}$ be 
 the azimuthal angle of $\overrightarrow{\theta_{E}}$ 
 in the laboratory coordinate system.
 In the case (a) $\eta\le\Theta_{Lab}^{}$,
 $\Phi_{E}$ varies within the region of 
 $(\Phi_{E}^{\max}-\Phi_{E}^{\min})\le\pi$.
 On the other hand,
 in the case (b) $\eta\ge\Theta_{Lab}^{}$,
 $\Phi_{E}$ varies within the whole region.
 Therefore we may expect that 
 some typical differences exist
 between the case (a) and (b)
 in the angular distribution for a process,
 for example $e^-e^+\to\gamma\gamma$,
 which are affected by the space-time noncommutativity.
%

 The magnetic-like component $\overrightarrow{\theta_{B}}$
 is also parametrized by the same way. 
 In general, however, both the direction and the magnitude
 of $\overrightarrow{\theta_{B}}$ are different from
 those of $\overrightarrow{\theta_{E}}$.
 Therefore $\theta_{\mu\nu}$ can be parameterized by 
 six parameters,
 four angles and two magnitudes of $\overrightarrow{\theta_{E}}$
 and $\overrightarrow{\theta_{B}}$,
 $\theta_E^{}=|\overrightarrow{\theta_{E}}|$ and 
 $\theta_B^{}=|\overrightarrow{\theta_{B}}|$.
 In the primary coordinate system,
\begin{eqnarray}
  \overrightarrow{\theta_E}=\theta_E
  \left( \overrightarrow{e_X} \sin\eta_E\cos\xi_E 
        +\overrightarrow{e_Y} \sin\eta_E\sin\xi_E 
        +\overrightarrow{e_Z} \cos\eta_E 
  \right),\\
  \overrightarrow{\theta_B}=\theta_B
  \left( \overrightarrow{e_X} \sin\eta_B\cos\xi_B 
        +\overrightarrow{e_Y} \sin\eta_B\sin\xi_B 
        +\overrightarrow{e_Z} \cos\eta_B 
  \right).
\end{eqnarray}
 By using (\ref{eqn:basis}),
 we can obtain the expression of 
 $\overrightarrow{\theta_{E}}$ and 
 $\overrightarrow{\theta_{B}}$ in 
 the laboratory coordinate system.
 We should take $\theta_E$ and $\theta_B$ as model parameters
 rather than the energy scale 
 $\Lambda_E=1/\sqrt{\theta_E}$ and
 $\Lambda_B=1/\sqrt{\theta_B}$.

\section{$e^-e^+\to \gamma\gamma$ in NCQED}
 A field theory in noncommutative space-time
 can be described equivalently
 by a field theory with commutative space-time variables
 in which every products of fields are replaced with
 the star product of fields.
 The star product is defined by
\begin{eqnarray}
 f\star{g(x)}=\left.
    \exp\left(\frac{i}{2}\del^{\mu}_{y}\theta_{\mu\nu}\del^{\nu}_{z}\right)
    f(y)g(z)\right|_{y=z=x},
\end{eqnarray}
 where $x$, $y$ and $z$ are ordinary commutative variables.

 NCQED action\cite{NCQED} is given by
\begin{eqnarray}
 S=\int d^4x\left(
       -\frac{1}{4}F^{\mu\nu}\star{F_{\mu\nu}}
       +i\bar{\Psi}\gamma^{\mu}\star{D_{\mu}\Psi}
       -m\bar\Psi\Psi\right)
\label{eqn:Sncqed}
\end{eqnarray}
 where 
 $F^{\mu\nu}=
   \del^{\mu}A^{\nu}-\del^{\nu}A^{\mu}-ie
 \left(A^{\mu}\star{A^{\nu}}-A^{\nu}\star{A^{\mu}}\right)$. 
 And covariant derivative for the matter fields is given by
 $D_{\mu}\Psi=\del_{\mu}\Psi-ieA_{\mu}\star\Psi$.
 We need nonlinear terms in field strength $F_{\mu\nu}$
 to keep NCQED action invariant under 
 noncommutative $U(1)_\star$ gauge transformation, 
\begin{eqnarray}
  A_{\mu}&\to& A_{\mu}'=U(x)\star{A_{\mu}}\star{U^{-1}(x)}
                   -\frac{i}{e}(\del_{\mu}U(x))\star{U^{-1}(x)},\\
  \Psi(x)&\to& \Psi'(x)=U(x)\star\Psi(x) ,\\
  \bar{\Psi}(x)&\to& \bar{\Psi}'(x)=\bar{\Psi}(x)\star{U^{-1}(x)}
\end{eqnarray}
 where 
\begin{eqnarray}
  U(x)=\exp(i\alpha(x))_\star\equiv\sum_{n=0}\frac{(i\alpha(x)\star)^n}{n!}
\end{eqnarray}
 and $U(x)\star{U}^{-1}(x)=U^{-1}(x)\star{U}(x)=1$.
 The NCQED action (\ref{eqn:Sncqed}) is invariant under
 the $U(1)_\star$ gauge transformation.

 In NCQED, we consider the pair annihilation process
 $e^-(k_1^{})e^+(k_2^{})\to\gamma(p^{}_1)\gamma(p^{}_2)$ 
 at future $e^-e^+$ linear colliders. 
 Each momentum is taken to be
\begin{eqnarray}
  \begin{array}{cccc}
\displaystyle
  k_1^{\mu}=\frac{\sqrt{s}}{2}(1, 0, 0, 1),&
\displaystyle
  k_2^{\mu}=\frac{\sqrt{s}}{2}(1, 0, 0,-1),&
\displaystyle
  p_1^{\mu}=\Bigl(\frac{\sqrt{s}}{2}, \overrightarrow{p}\Bigr),&
\displaystyle
  p_2^{\mu}=\Bigl(\frac{\sqrt{s}}{2}, -\overrightarrow{p}\Bigr),\cr
  \end{array}
\end{eqnarray}
 where $\overrightarrow{p}=
 (\sqrt{s}/2)(s_\theta^{}c_\phi^{}, 
              s_\theta^{}s_\phi^{}, 
              c_\theta^{})$.
 $\theta$ and $\phi$ are polar angle and azimuthal angle
 of final state photon
 in the laboratory coordinate system.
 The differential cross section for
 $e^-(k_1^{})e^+(k_2^{})\to\gamma(p^{}_1)\gamma(p^{}_2)$
 in the center of mass system is given by
\begin{eqnarray}
  \frac{d\sigma}{d\cos\theta d\phi}
&=&
  \frac{\alpha^2}{4s}\left(\frac{t}{u}+\frac{u}{t}
            -4\frac{t^2+u^2}{s^2}\sin^2\Delta_{NC}\right),
\label{eqn:dcross} \\
 \Delta_{NC}&=&\frac{p^{\mu}_1\theta_{\mu\nu}p^{\nu}_2}{2}
    =-\left(\frac{s}{4}\right)
    \frac{\overrightarrow{\theta_E}\cdot \overrightarrow{p}}
    {|\overrightarrow{p}|},
\label{eqn:delta}
\end{eqnarray}
 where 
 $s, t$ and $u$ are usual Mandelstam variables,
 $s=(k_1^{}+k_2^{})^2$, $t=(k_1^{}-p_1^{})^2$
 and $u=(k_1^{}-p_2^{})^2$.
 The $\overrightarrow{\theta_E^{}}$ is given 
 in (\ref{eqn:thE}), (\ref{eqn:thEV}) and (\ref{eqn:thES}).
 When $\Delta_{NC}=0$, 
 differential cross section (\ref{eqn:dcross}) corresponds to 
 the differential cross section in QED.

 Since two photons in the final state are identical,
 we cannot distinguish two configuration 
 $(\theta,\phi)$ and $(\pi-\theta, \pi+\phi)$.
 Therefore,
 we must get the sum of the differential cross section 
 for $(\theta,\phi)$ and $(\pi-\theta, \pi+\phi)$.
 Then the observable is 
\begin{eqnarray}
  \frac{d\sigma_{obs}^{}}{d\cos\theta d\phi}&=&
   \frac{d\sigma}{d\cos\theta d\phi}(\theta,\phi)  
  +\frac{d\sigma}{d\cos\theta d\phi}(\pi-\theta,\pi+\phi).
\label{eqn:sigobs}
\end{eqnarray}
 Note that (\ref{eqn:sigobs})
 is defined in the region 
 $0\le \cos\theta< 1$ and $0\le\phi\le2\pi$.
 It is easy to show from (\ref{eqn:dcross})
 and (\ref{eqn:delta}) that
 $\Delta_{NC}(\pi-\theta,\pi+\phi)=$
 $-\Delta_{NC}(\theta,\phi)$ and 
\begin{eqnarray}
   \frac{d\sigma}{d\cos\theta d\phi}(\Delta_{NC}^{})  
  =\frac{d\sigma}{d\cos\theta d\phi}(-\Delta_{NC}^{}).
\end{eqnarray}
 Moreover, 
 this imply that the differential cross section
 (\ref{eqn:dcross}) and (\ref{eqn:sigobs})
 are symmetric for
 the change of the sign of $\overrightarrow{\theta_{E}^{}}$,
 $\overrightarrow{\theta_E^{}}\leftrightarrow
   -\overrightarrow{\theta_E^{}}$.
 Therefore we cannot distinguish
 between $(\eta, \xi)$ and $(\pi-\eta, \pi+\xi)$
 by observing the process $e^-e^+\to\gamma\gamma$.
 There is two-fold ambiguity
 for the determination of $(\eta, \xi)$. 
 
 We can see from (\ref{eqn:dcross}) that
 NCQED effect to the differential cross section of
 $e^-e^+\to \gamma\gamma$ always gives
 the negative contribution, 
 moreover, from (\ref{eqn:delta}) we find
\begin{eqnarray}
 \begin{array}{crc}
  |\Delta_{NC}|=\max(\Delta_{NC})
           =\displaystyle\frac{\, s\, }{4}\theta_E
&{\rm if}& \overrightarrow{p}\parallel\overrightarrow{\theta_E},\\
   \Delta_{NC}=0            
&{\rm if}& \overrightarrow{p}\perp \overrightarrow{\theta_E}.
 \end{array}
\end{eqnarray}
 This means that,
 when we compare NCQED prediction with QED prediction,
 the deficit of the differential cross section 
 appears around the specific direction in which
 $\overrightarrow{p}$ is almost parallel to
 $\overrightarrow{\theta_E}$.
 Furthermore, such a specific direction varies with time
 in the laboratory coordinate system,
 as we have discussed in previous section.
 Therefore, in general, 
 observables for $e^-e^+\to\gamma\gamma$
 in the laboratory coordinate system have
 time dependence even for the total cross section.

 We may consider that
 the measured value for observable 
 by collider experiments is usually
 given as a mean value. And
 such a mean value should be compared with 
 NCQED prediction averaged over time.
 Taking into consideration that
 the period of time variation of the observables in NCQED 
 is  the sidereal day $T_{day}$,
 we introduce the time averaged observables 
 as follows;
\begin{eqnarray}
  \left\langle\frac{d{\sigma}}{d\cos\theta{d\phi}}\right\rangle_{T}
  &\equiv&
  \frac{1}{T_{day}}
  \int^{T_{day}}_{0}\frac{d\sigma_{obs}^{}}{d\cos\theta{d\phi}}dt,
\\
  \left\langle\frac{d{\sigma}}{d\cos\theta}\right\rangle_{T}
  &\equiv&
  \frac{1}{T_{day}}
      \int^{T_{day}}_{0}\frac{d\sigma_{obs}^{}}{d\cos\theta}dt,
\\
  \left\langle\frac{d{\sigma}}{d\phi}\right\rangle_{T}
  &\equiv&
  \frac{1}{T_{day}}\int^{T_{day}}_{0}\frac{d\sigma_{obs}^{}}{d\phi}dt,
\label{eqn:dcr_dphi}
\\
  \langle\sigma\rangle_{T} &\equiv&
  \frac{1}{T_{day}}\int^{T_{day}}_{0}\sigma_{obs}^{}dt,
\end{eqnarray}
where
\begin{eqnarray}
\frac{d{\sigma}_{obs}^{}}{d\cos\theta}&\equiv&
  \int^{2\pi}_{0}\!\!d\phi
      \frac{d\sigma_{obs}^{}}{d\cos\theta{d\phi}},\\
\frac{d{\sigma}_{obs}^{}}{d\phi}&\equiv&
  \int^{1-\epsilon}_{0}\!\!d(\cos\theta)
      \frac{d\sigma_{obs}^{}}{d\cos\theta{d\phi}},\\
{\sigma}_{obs}^{}&\equiv&
  \int^{1-\epsilon}_{0}\!\!d(\cos\theta)
  \int^{2\pi}_{0}\!\!d\phi
      \frac{d\sigma_{obs}^{}}{d\cos\theta{d\phi}}.
\end{eqnarray}
 The polar angle cut is denoted by
 $\epsilon\,$ ($0\le\epsilon\le1$). 
 It is easy to see that 
 $\langle\sigma\rangle_T=\sigma_{obs}^{}$
 which is usual total cross section,
 when $\sigma_{obs}^{}$ is independent of time.

 Note that we have integrated out
 the $\xi$ dependence of the observables
 by taking average over time,
 since $\xi$ play a role of initial phase for time evolution.
 Therefore 
 $\theta_E^{}$ and angle $\eta$
 may be determined by 
 the time averaged observables.

\section{ Numerical Results }
 We show several characteristic results in NCQED
 and also discuss how to prob $\overrightarrow{\theta_E^{}}$
 by using observables in the laboratory coordinate system.
 We set the laboratory coordinate system
 by taking ($\delta$, $a$)=($\pi/4$, $\pi/4$).
 The cut for $\cos\theta$ is taken $\epsilon=0.2$.

\subsection{Azimuthal angle distribution}
\begin{figure}[tb]
\begin{center}
\begin{minipage}[c]{150mm}
\centerline{
\includegraphics[width=80mm, angle=-90]{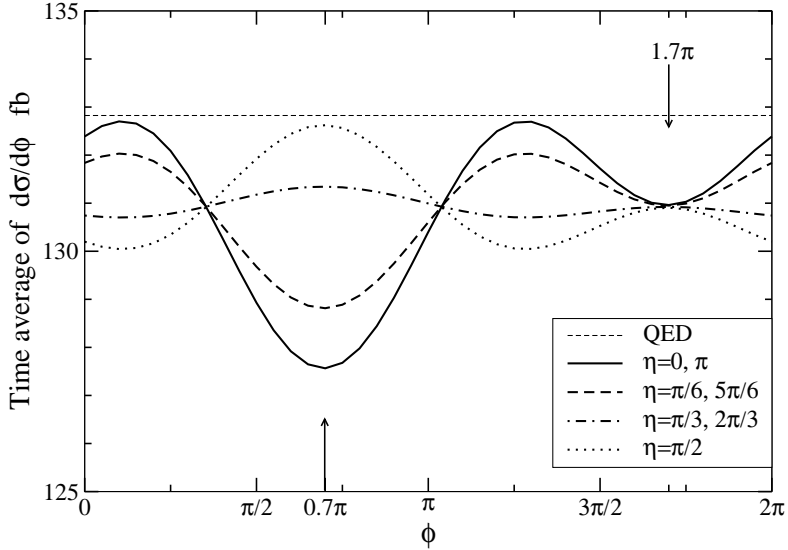}
}
\caption[]{Time averaged azimuthal angle distribution
   for $\eta=0$, $\pi/6$, $\pi/3$, $\pi/2$,
   $2\pi/3$, $5\pi/6$, $\pi$.
   We set the laboratory coordinate system by taking
 $(\delta, a)=(\pi/4, \pi/4)$. 
   We take $\sqrt{s}=500$GeV and 
   $\theta_E^{}=(500 {\rm GeV})^{-2}$.
}
\label{fig:dcrbar_phi}
\end{minipage}
\end{center}
\end{figure}

 Anisotropy of azimuthal angle distribution 
 of $e^-e^+\to\gamma\gamma$ is predicted in NCQED
 even if we consider the time averaged distribution
 $\langle{d\sigma/d\phi}\rangle_{T}^{}$.
 Figure \ref{fig:dcrbar_phi} shows 
 $\langle{d\sigma/d\phi}\rangle_{T}^{}$
 for  $\theta_E^{}=(500\mbox{GeV})^{-2}$
 and several values of $\eta$. 
 We take $\sqrt{s}=500$GeV.

 We see from figure \ref{fig:dcrbar_phi}
 that the curves of 
 $\langle{d\sigma}/d\phi\rangle_T^{}$
 are sensitive to 
 the value of $\eta$ around $\phi\simeq 0.7\pi$
 and also almost 
 independent of the value of $\eta$
 around $\phi\simeq 1.7\pi$.
 Furthermore
 $\langle{d\sigma/d\phi}\rangle_{T}^{}$
 is almost flat
 around $\phi\simeq 1.7\pi$
 for any $\eta$.

 Those specific angles $0.7\pi$ and $1.7\pi$
 can be interpreted as 
 the azimuthal angle of
 $\overrightarrow{\theta_{ES}^{}}$
 and $-\overrightarrow{\theta_{ES}^{}}$
 in the laboratory coordinate system.
 In other words,
 those are the azimuthal angles of 
 north pole and south pole of the celestial sphere.
 Since we take $(\delta$, $a)=(\pi/4,$ $\pi/4)$, 
 the azimuthal angle $\phi_{ES}^{N}$
 of $\overrightarrow{\theta_{ES}^{}}$
 can be derived from (\ref{eqn:thES}) 
 as follows
\begin{eqnarray}
  \cos\phi_{ES}^{N}
    =\frac{-c_\delta^{}s_{a}^{}}{\sqrt{1-c^2_{\delta}c^2_{a}}}
    =-\frac{1}{\sqrt{3}}
&,&
  \sin\phi_{ES}^{N}
    =\frac{s_\delta^{}}{\sqrt{1-c^2_{\delta}c^2_{a}}}
    =\sqrt{\frac{\,2\,}{\,3\,}}
\end{eqnarray}
 then $\phi_{ES}^{N}\simeq 0.7\pi$ and
 the azimuthal angle of $-\overrightarrow{\theta_{ES}^{}}$
 is given by 
 $\phi_{ES}^{N}+\pi\simeq 1.7\pi$. 

 We also see from figure  \ref{fig:dcrbar_phi}
 that each input $\eta$ and $\pi-\eta$
 gives the same distribution of
 $\langle{d\sigma}/d\phi\rangle_T^{}$. 
 This is because
 the differential cross section of $e^-e^+\to\gamma\gamma$
 is symmetric for 
 $\overrightarrow{\theta_E^{}} \leftrightarrow
   -\overrightarrow{\theta_E^{}}$.

 We may determine $\eta$,
 except for two-fold ambiguity for $\eta$ and $\pi-\eta$, 
 by fitting the shape of curve
 of $\langle{d\sigma/d\phi}\rangle_T^{}$,
 especially around $\phi\simeq \phi_{ES}^{N}=0.7\pi$.
 Also we may determine $\theta_E^{}$ 
 almost independently of $\eta$
 by measuring the deficit 
 of $\langle{d\sigma/d\phi}\rangle_T^{}$ compared with 
 QED prediction around $\phi\simeq 1.7\pi$.

\subsection{Time dependent total cross section}
 In order to determine $\xi$, we need to trace
 the apparent time variation of observables
 due to the earth's rotation.
 Since total cross section $\sigma_{obs}^{}$
 depends on $\theta_E^{}$, $\eta$ and also $\xi$,
 we may expect that 
 $\theta_E^{}$, $\eta$ and $\xi$
 could be determined
 by measuring time variation of $\sigma_{obs}^{}$ precisely,
 except for the two-fold ambiguity for ($\eta$, $\xi$).

\begin{figure}[tb]
\begin{center}
\begin{minipage}[c]{150mm}
\centerline{
\includegraphics[width=80mm, angle=-90]{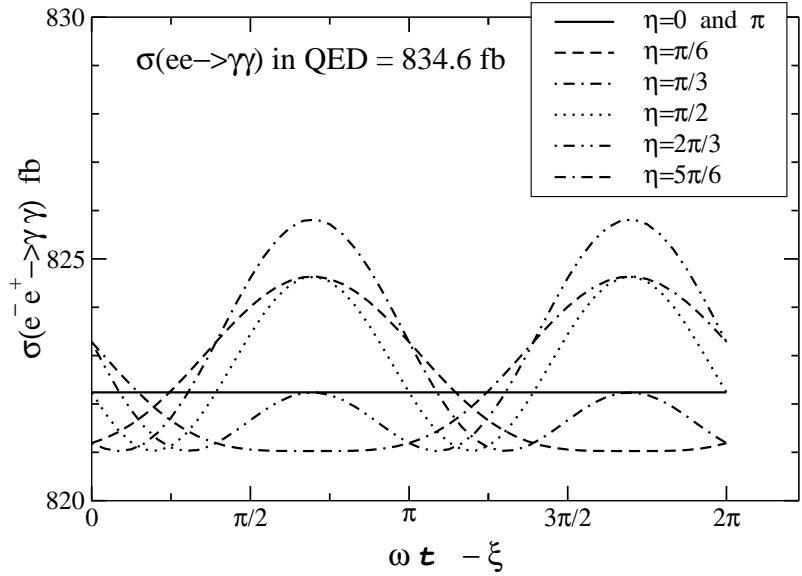}
}
\caption[]{Apparent time variation of total cross section 
 for  $\eta=0,\pi/6,\pi/3,\pi/2,2\pi/3,5\pi/6$ and $\pi$.
 We take $(\delta,a)=(\pi/4, \pi/4)$, $\sqrt{s}=500$GeV and 
 $\theta_{E}=(500$GeV$)^{-2}$.
     }
\label{fig:crtime}
\end{minipage}
\end{center}
\end{figure}
\begin{figure}[tb]
\begin{center}
\begin{minipage}[c]{150mm}
\centerline{
\includegraphics[width=80mm, angle=-90]{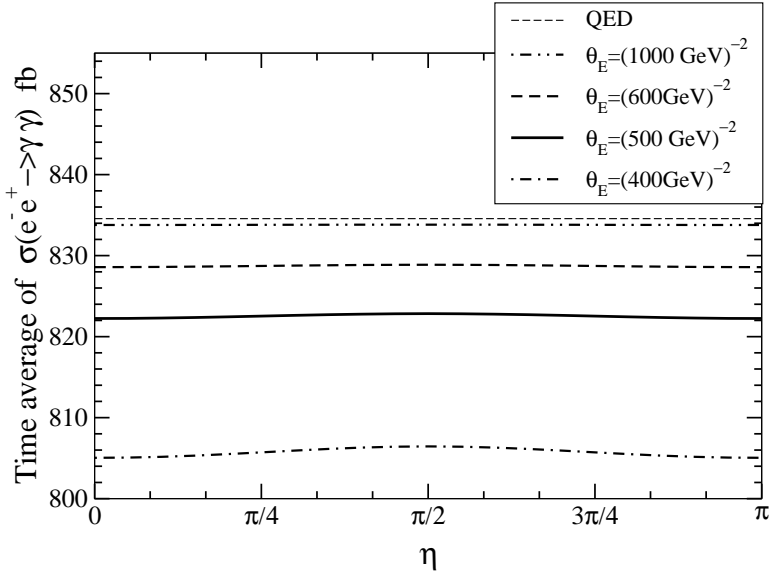}
}
\caption[]{Time averaged total cross section
   of $e^-e^+\to\gamma\gamma$ 
   for ${\theta_{E}}=(400{\rm GeV})^{-2},$ 
   $(500{\rm GeV})^{-2},$ $(600{\rm GeV})^{-2}$
   and $(1000{\rm GeV})^{-2}$ are shown.
   Horizontal axis is taken to be angle $\eta$.
   QED result is also shown.
   We take $(\delta, a)=(\pi/4, \pi/4)$ and $\sqrt{s}=500$GeV.
   }
\label{fig:crbar}
\end{minipage}
\end{center}
\end{figure}

 Figure \ref{fig:crtime} shows $\sigma_{obs}^{}$
 as a function of $\omega{t}-\xi$
 for $\theta_E^{}=(500\mbox{GeV})^{-2}$
 and for several values of $\eta$.
 We can see from figure \ref{fig:crtime} that
 the shape of curve is sensitive to the $\eta$.
 If the time variation of total cross section is observed,
 we could determine both magnitude and direction of
 $\overrightarrow{\theta_E^{}}$ by
 fitting the NCQED prediction of $\sigma_{obs}^{}$
 with the data in three parameters space
 ($\theta_E^{},\eta,\xi$).
 The magnitude $\theta_E^{}$ and the angle $\eta$ 
 may be determined by the fitting of both
 the magnitude and the shape of curves of $\sigma_{obs}^{}$.
 The $\xi$ may be determined by
 the measurement of the phase of time evolution
 of $\sigma_{obs}^{}$.
 However, as is mentioned previously,
 we cannot distinguish 
 ($\eta$, $\xi$) from ($\pi-\eta$, $\pi+\xi$).
 For example,
 the graph of $\sigma_{obs}^{}$ for $\eta=\pi/3$
 is identical to that for $\eta=2\pi/3$
 shifted the phase $\xi$ to $\xi+\pi$.
 
 Although we may determine $\overrightarrow{\theta_E}$
 by tracing the time variation of
 the differential cross section of $e^-e^+\to\gamma\gamma$
 instead of total cross section,
 we can easily imagine that such an experiment
 needs very large luminosity.
 This is because we must divide not only the phase space
 but also the time distribution into many bins,
 in order to trace the time variation.
 Therefore, in the determination of 
 $\theta_E^{}$, $\eta$ and $\xi$,
 we had better probe the time variation of
 total cross section in the early experiments
 at $e^-e^+$ linear colliders.

\subsection{$\langle{d\sigma/d\phi}\rangle_T^{}$ vs. $\sigma_{obs}$}
 We can see from figures \ref{fig:dcrbar_phi} and \ref{fig:crtime} 
 that $\langle{d\sigma/d\phi}\rangle_T^{}$ and $\sigma_{obs}$
 show opposite behavior for each input value of $\eta$.
 For example,
 when $\eta=0 \mbox{ or } \pi$,
 we may observe large variation of
 azimuthal angle distribution  
 $\langle{d\sigma/d\phi}\rangle_T^{}$.
 In this case,
 we find no time variation of $\sigma_{obs}$.
 On the contrary,
 when $\eta=\pi/3 \mbox{ or } 2\pi/3$,
 since the variation of
 $\langle{d\sigma/d\phi}\rangle_T^{}$
 is very small,
 we may observe the flat distribution in the experiments.
 In this case,
 we find large time variation of $\sigma_{obs}$.
 Therefore we may expect that
 non-uniform distribution due to NCQED effect
 should appear in the 
 $\langle{d\sigma/d\phi}\rangle_T^{}$
 and/or $\sigma_{obs}$ for any value of $\eta$,
 if ${\theta_E^{}}$ is large enough.

 \subsection{Time averaged total cross section}
 Finally, we consider what we can measure
 by the time averaged total cross section
 $\langle\sigma\rangle_T$.
 Figure \ref{fig:crbar} shows $\langle\sigma\rangle_T$
 as a function of $\eta$.
 It is easy to see that  $\langle\sigma\rangle_T^{}$
 is almost independent of $\eta$.
 In case of us observing
 some deficit of $\langle\sigma\rangle_T^{}$,
 we may determine $\theta_E^{}$
 independently of $\eta$
 by measuring $\langle\sigma\rangle_T^{}$. 
 
 On the other hand,
 in case of us observing no signal,
 we may obtain the upper limit on $\theta_E^{}$.
 The $1\sigma$ deviation for 
 total cross section in QED, $\sigma_{QED}$,
 can be estimated by $\sqrt{{\sigma_{QED}}/{L}}$. 
 For $\sqrt{s}=500$GeV and $\epsilon=0.2$,
 we have
 $({28.89}/{\sqrt{L}})$
 where $L$ is the luminosity given in fb$^{-1}$.
 In this case,
 an expected 95\%CL upper limit on $\theta_E^{}$ 
 is found to be
\begin{eqnarray}
 \begin{array}{cl}
  \theta_E^{}\lsim (600{\rm GeV})^{-2} & \mbox{for $L$=100fb$^{-1}$}.
 \end{array}
\label{eqn:L100}
\end{eqnarray}
 Furthermore, since
 $|\sigma_{QED}-\langle\sigma\rangle_T^{}|
            \propto{(s\theta_E)}^2$ for $|s\theta_E|<1$,
 we may estimate
 95\%CL upper limit on $\theta_E^{}$ for arbitrary $L$
 from ($\ref{eqn:L100}$) as follows;
\begin{eqnarray}
  \theta_E^{}\lsim (600{\rm GeV})^{-2}
  \left(\frac{100 {\rm fb^{-1}}}{L}\right)^{1/4}.
\end{eqnarray}
For example, we find
$\theta_E^{}\lsim (800{\rm GeV})^{-2}$
when $L$=1000fb$^{-1}$.

\section{Conclusion and Remarks}
 We have presented phenomenological formulation of
 the apparent time variation of noncommutativity parameter
 $\theta_{\mu\nu}$ in the laboratory coordinate system.
 In our framework, the laboratory coordinate system
 have been taken to be a familiar coordinates 
 to the collider experiments,
 and the primary coordinate system 
 fixed to the celestial sphere have been introduced.
 We have shown the transformation formula
 between the primary and the laboratory coordinate system,
 and also shown the expression of
 $\overrightarrow{\theta_E^{}}$
 in the laboratory coordinate system.
 The formulation presented in this paper
 is applicable to the study of models
 which predict an intrinsic direction of
 the space-time\cite{NCQED,sugamoto}.

 As an example,
 we have applied our formulation to NCQED
 and discussed the determination of
 $\overrightarrow{\theta_E^{}}$
 at the $e^-e^+$ linear collider experiments
 by the process $e^-e^+\to\gamma\gamma$.
 The
 $\overrightarrow{\theta_E^{}}$
 have been parameterized by  $\theta_E^{}$, $\eta$ and $\xi$
 in the primary coordinate system.
 We have shown that
  $\overrightarrow{\theta_E^{}}$ may be determined
 by the detailed study of
 the time dependent total cross section,
 though two-fold ambiguity in the parameter space
 ($\eta$, $\xi$) remains.
 To determine $\xi$, we need to probe the phase
 of the time evolution of $\sigma_{obs}$.
 In case of us observing no signal,
 probably this is the most realistic case,
 we may obtain the upper limit on $\theta_E^{}$
 independently of the direction of
 $\overrightarrow{\theta_E^{}}$.

 So far we have considered
 one experiment with ($\delta$, $a$)=($\pi/4$, $\pi/4$).
 If there are several detector sites in
 the $e^-e^+$ collider experiment and 
 the direction of $e^-$ beam in each site
 is set to be along to the different direction,
 such as four LEP experiments, then 
 the angular distributions of $e^-e^+\to\gamma\gamma$
 and the time variation of observables
 should behave differently in each experiment.
 This is because the direction of 
 $\overrightarrow{\theta_{ES}^{}}$
 in the laboratory coordinate system at 
 one detector site differ from that at other detector sites.
 Therefore we can expect that
 the combined analysis of several experiments
 with the different ($\delta$, $a$)
 play an important role in the attempt 
 to probe the space-time noncommutativity.

 Finally,
 we would like to make some comments on
 the determination of the magnetic-like component
 $\overrightarrow{\theta_B^{}}$.
 Since the process $e^-e^+\to\gamma\gamma$
 is independent of $\overrightarrow{\theta_B^{}}$,
 to determine $\overrightarrow{\theta_B^{}}$,
 we must consider other processes relevant to
 $\overrightarrow{\theta_B^{}}$,
 for example $e^-\gamma\to{e^-}\gamma$ process 
 which depend on both $\overrightarrow{\theta_E^{}}$
 and $\overrightarrow{\theta_B^{}}$.
 The process $\gamma\gamma\to\gamma\gamma$ 
 may also be available
 to determine $\overrightarrow{\theta_B^{}}$.
 By combining the results from those processes,
 we may determine $\overrightarrow{\theta_E^{}}$
 and $\overrightarrow{\theta_B^{}}$.
 We postpone the study of this matter to the future studies.

 \medskip
 \noindent{\it Acknowledgments:}
 The author thanks
 O. Kamei and A. Sugamoto
 for reading the manuscript and useful comments,
 and
 K. Hagiwara and T. Kawamoto 
 for useful discussion and comments.
 The author also thanks I. Watanabe for
 useful comments on
 the usefulness of $X$ axis defined by 
 the vernal equinox $\Upsilon_{2000.0}$.

\end{document}